\documentclass[intlimits,twoside,a4paper]{article}

\usepackage{amsmath,amssymb}
\usepackage{graphicx}

\usepackage{algorithm}
\usepackage{algorithmic}

\usepackage[T2A]{fontenc}
\usepackage[cp1251]{inputenc}

\usepackage[eqsecnum]{cmpj2}

\issue{2014}{17}{3}{33801}
\doinumber{10.5488/CMP.17.33801}

\title[Collective behavior with chat Bots]{Collective emotion dynamics in chats with  agents, moderators and  Bots}
\author{M. \v{S}uvakov\refaddr{label1,label2}, B. Tadi\'c\refaddr{label1}
}
\addresses{
\addr{label1} Department of Theoretical Physics, Jo\v{z}ef Stefan Institute, Jamova 39, 1001 Ljubljana, Slovenia
\addr{label2} Institute of Physics Belgrade, University of Belgrade, Pregrevica 118, 11080 Beograd, Serbia
}

\authorcopyright{M. \v{S}uvakov, B. Tadi\'c, 2014}

\date{Received March 27, 2014, in final form April 19, 2014}

\begin{document}

\maketitle

\begin{abstract}
Using agent-directed simulations, we investigate fluctuations in the collective emotional states on a chat network where agents interchange  messages  with a  fixed number of moderators and emotional Bot. To design a realistic chat system, the interaction rules and some statistical parameters, as well as the agent's attributes, are inferred from the empirical chat channel \texttt{Ubuntu}. In the simulations,  the Bot's emotion is fixed;  the moderators tune the level of its activity by passing a fraction $\epsilon$ of messages to the Bot. At $\epsilon \gtrsim 0$, the collective emotional state matching the Bot's emotion polarity gradually arises; the average growth rate of the dominant emotional charge serves as an order parameter.   Due to self-organizing effects, the collective dynamics  is more explosive when positive emotions arise by positive Bot than the onset of negative emotions in the presence of  negative Bot at the same $\epsilon$.
Furthermore, when the emotions matching the Bot's emotion polarity are spread over the system, the underlying fractal processes exhibit higher persistence and stronger clustering of events than the processes spreading of emotion polarity opposite to the Bot's emotion. On the other hand, the relaxation dynamics is controlled by the external noise;  the related nonextensive parameter,  estimated from the statistics of returns, is virtually independent of the Bot's activity level and emotion contents.
\pacs 89.20.-a, 05.10.-a, 05.65.+b, 64.60.aq, 89.75.Fb

\end{abstract}

\section{Introduction\label{sec-intro}}
In recent years, application of statistical physics into interdisciplinary research of social dynamics becomes of key importance, especially  for understanding the fundamental problem --- how the collective behaviors emerge over time from the activity of individual participants. Owing to the wealth of empirical data from various online communication systems,  social dynamics has been researched as an example of a complex system in the physics laboratory \cite{mitrovic2010a,mitrovic2010c,ST12,ST_Plos2013,dodds2011,crowd-search,we-MySpace11,we-Chats1s,warsaw2011}. In these studies, methods of statistical physics and graph theory are often supplemented with the machine-learning techniques for text analysis to infer particular \textit{content}, i.e., information or emotion \cite{emotions-Oxford2007,paltoglouIEEE2011,skowron-paltoglou2011afect}  communicated between users. Having high temporal resolution of the data, the role of emotions  was investigated on  Blogs  and Diggs \cite{mitrovic2010a,mitrovic2011}, in online games \cite{ST12}, forums \cite{warsaw2011}, online social networks \cite{we-MySpace11,facebook2012}, crowd-sourcing \cite{crowd-search}, and others. It has been recognized that  emotions play a role in shaping the structure of social networks that evolve with user  interactions \cite{we-entropy,we-Chats-chapter}; moreover,  the emergence of  patterns of positive and negative conduct \cite{ST_Plos2013}, as well as collective  emotional behaviors \cite{collective_emo_book} have been studied.

Theoretical modelling of the underlying stochastic processes, however,  is much more complex in comparison with the processes in standard physical systems of interacting units (atoms, molecules, nanoparticles). Specifically, interactions among human users essentially depend on their circadian cycles and a number of user-specific characteristics, such as its psychology profile, activity pattern, mutual trust or preferences towards particular communication contents. Recent advances in the computational science \cite{sloot2012} with individual-based modelling  provide the right platform where these attributes of humans can be effectively considered \cite{BT_ABMbook}. In this approach, that we also use here, the empirical data of a concrete online system are  analysed to infer the activity patterns and statistical distributions that characterize the users' behavior. Then, these distributions are utilized to design a set of attributes of  interacting agents  \cite{we_ABMblogs12,we-abm_ExpBots,we-ABMrobots,BT_ABMbook} (see section~\ref{sec-model}).

In this paper, we use agent-directed simulations to investigate the  collective dynamics with emotions in online chat systems. To design a realistic chat system, we adopt rules and parameters of the empirical channel  \texttt{Ubuntu}  \cite{Ubuntu-story}; in the system, an automated chat program (chat Bot) is present as well as a number of moderators (users who manage new messages posted on the channel passing them to the Bot). In the model, agents and moderators are  interacting by  interchange of messages with each other and with an emotional chat Bot whose emotion is fixed during the process. The emotional state of each participant in the process is described by two variables, representing two key elements of emotion: arousal (degree of reactivity) and valence (pleasure or displeasure).  Quantitative study of emotions in psychology is based on Russell's model of affect \cite{russell1980}, where these two components are sufficient to recognize a commonly known emotion. Conventional values of these quantities are expressed by  mapping onto the surface of a unit circle. For consistency, several  agent's attributes are taken from the distributions, which are obtained from the same empirical system. A description of the empirical data, the structure of chat-networks and the role of emotional arousal inferred from the analysed text messages, can be found in works \cite{we-Chats-conference,we-Chats-chapter,we-Chats1s}. By analogy with the empirical system, in the simulations, the chat network co-evolves with the dynamics, i.e., by adding nodes and links or increasing  weight on previous links \cite{we-entropy}. Each node of the network represents an agent, moderator or Bot, and a directed link indicates that at least one  message from the source to target node  occurred during the evolution time. The emergent  ``social'' contacts facilitate further communications; hence, spreading of emotion among the connected agents occurs in a self-organized manner. In stochastic processes on networks, robust scaling features are  often observed where both nature of the process and structure of the network play a role \cite{tadic2007,tadic2002}.   Here, we focus on the collective dynamics of emotion spreading on chat  networks.

Recently,  we have simulated the presence of Bots with human-like features (personalized emotional messages, delayed actions) in chat systems \cite{we-abm_ExpBots,we-ABMrobots}. It was demonstrated that, despite a limited number of contacts,  the Bot's activity can affect the mood of the agents in the entire system.  In contrast to works of \cite{we-abm_ExpBots,we-ABMrobots}, here we introduce a variant of the agent-based model in which the Bot's activity level can be tuned by varying a parameter, which helps us investigate how  the collective emotional state of agents is raised by Bots. The control parameter is the fraction of messages passed to the Bot by the moderators. Apart from potentials of the model to predict the situations at the real chat channel, the simulated emotion dynamics represents a theoretically interesting case of an open  nonlinear system.  The system is driven from outside by the arrival of new agents according to the empirical pink noise signal, as well as from inside, by a sequence of the Bot's  messages carrying a fixed emotion content. Analysing the simulated data, we show how the collective emotional states of agents arise under the influence of Bots. Moreover, we  demonstrate that the fractal structure of time series and clustering of events with a specified emotion content have universal features connected with $q$-generalized statistics.

In section~\ref{sec-model}, we introduce the model by defining the agent's profiles, the parameters and  interaction rules. Some features of the emergent network and quantities measuring the collective emotional states are studied in section~\ref{sec-response}. Then, we investigate the universal dynamics of emotional chats in section~\ref{sec-tseries}. A brief summary is given in section~\ref{sec-conclusion}.

\section{Agent-based model of emotional chats with moderators and Bots\label{sec-model}}

Agents are situated on an evolving chat network. In a network, the appearance of a directed link $i\to j$ at time $t$ indicates that, at that time, a message was sent  from an agent $i$ to agent $j$. The network size $N(t)$ increases by the arrival of new agents; active agents establish new links or use previous connections. Hence, the connected agents can influence each other along the links that exist up to the current time. Each agent can see all messages appearing on the channel, including the messages that are not directed to anyone of the agents. These messages are exposed within the rolling time window, i.e., all messages created on the channel within the past $T_0$ steps are visible.  We assume that the messages  created within $T_0$ previous time steps and directed to an agent, as well as messages exposed on the channel, can influence the agent's emotional state. Their aggregated impact of the agent's $i$ arousal and valence is built via the corresponding  \textit{influence fields},  $h^{\mathrm{a}}_{i}(t)$ and $h^{\mathrm{v}}_{i}(t)$, respectively:
\begin{equation}
h^{\mathrm{a}}_{i}(t)=\frac{\sum_{j\in
    \mathrm{lin}_{i}}a^{m}_{j}\left\{\theta\left[t_{m}-(t-1)\right]-\theta\left[t_{m}-(t-T_{0})\right]\right\}}{\sum_{j\in
  \mathrm{lin}_{i}}\left\{\theta\left[t_{m}-(t-1)\right]-\theta\left[t_{m}-(t-T_{0})\right]\right\}} \, ,
\label{arousal_field}
\end{equation}
and
\begin{equation}
  h^{\mathrm{v}}_{i}(t)=\frac{1-0.4r_{i}(t)}{1.4}\frac{N^{\mathrm{p}}_{i}(t)}{N^{\mathrm{emo}}_{i}(t)}-\frac{1+0.4r_{i}(t)}{1.4}\frac{N^{\mathrm{n}}_{i}(t)}{N^{\mathrm{emo}}_{i}(t)}
\, . \label{valence_field}
\end{equation}
Here, $a^{m}_{j}$ is the arousal contained in the message $m$ along the link $j\to i$; the summation is over the messages from the list $\mathrm{lin}_{i}$ of agent's $i$ incoming links.  $N^{\mathrm{p}}_{i}$ and $N^{\mathrm{n}}_{i}$ are the number of the messages directed to $i$ within the considered time window which carry positive and negative emotion valence, respectively, and $N^{\mathrm{emo}}_{i}(t)=N^{\mathrm{p}}_{i}(t)+N^{\mathrm{n}}_{i}(t)$; $\theta(t)$ is Heaviside theta function and $r_i(t)=\mathrm{sgn}[v_i(t)]$ is the sign of an emotional valence of agent $i$ at time $t$.

The messages posted on the channel only equally affect each agent. They are taken into account by the corresponding components of the mean fields  $h^{\mathrm{a}}_{\mathrm{mf}}$ and $h^{\mathrm{v}}_{\mathrm{mf}}$. These fields are computed in a way similar to the expressions (\ref{arousal_field}) and (\ref{valence_field}), but the summation is over the set $\cal{S}$ of all messages, i.e.,
\begin{equation}
h^{\mathrm{a}}_{\mathrm{mf}}=\frac{\sum_{j\in
    \cal{S}}a^{m}_{j}\left\{\theta\left[t_{m}-(t-1)\right]-\theta\left[t_{m}-(t-T_{0})\right]\right\}}{\sum_{j\in
    \cal{S}}\left\{\theta\left[t_{m}-(t-1)\right]-\theta\left[t_{m}-(t-T_{0})\right]\right\}}
\, , \label{arousal_mfield}
\end{equation}
and similarly,  $ h^{\mathrm{v}}_{\mathrm{mf}}(t)$ contains contributions of these messages in the manner of equation~(\ref{valence_field}).

According to the line of modelling the agents' interaction in  networks \cite{we_ABMblogs12,we-abm_ExpBots,we-ABMrobots}, the agent's $i$ emotional arousal $a_i(t)$ and valence $v_i(t)$ fluctuate  under the action exerted on that agent over time. Specifically, when the agent $i$ is taking part in the process (according to the rules), the current values of its arousal and valence are corrected by the field's contribution \cite{we-ABMrobots}:
\begin{equation}
a_{i}(t+1)=
  (1-\gamma_{\mathrm{a}})a_i(t)+\left(\frac{h^{\mathrm{a}}_{i}(t)+kh^{\mathrm{a}}_{\mathrm{mf}}(t)}{1+k}\left\{1+d_{2}\left[a_{i}(t)-a_{i}(t)^{2}\right]\right\}
  \right)[1-a_{i}(t)],
  \label{arousal_map}
\end{equation}
\begin{equation}
v_{i}(t+1)=
  (1-\gamma_{\mathrm{v}})v_i(t) +\left(\frac{h^{\mathrm{v}}_{i}(t)+kh^{\mathrm{v}}_{\mathrm{mf}}(t)}{1+k}\left\{1+c_{2}\left[v_{i}(t)-v_{i}(t)^{3}\right]\right\}
  \right)[1-|v_{i}(t)|] ,
   \label{valence_map}
\end{equation}
$i=1,2,\dots, N(t)$.
Then, \textit{an elevated arousal may trigger the agent's action}, i.e., cretaing a new message which carries the agent's current emotion. The agent's activity status is regulated by the dynamic rules that are described in the following. Otherwise, when the agent is not involved in any activity, its arousal and valence are  constantly relaxing towards zero with the rates $\gamma_{\mathrm{a}}=\gamma_{\mathrm{v}}$. Note that  the maps in equations~(\ref{arousal_map}) and (\ref{valence_map})  systematically preserve the variables  in the range $a_i(t)\in[0,1]$ and $v_i(t)\in [-1,+1]$.

\begin{figure}[!ht]
\includegraphics[width=0.495\textwidth]{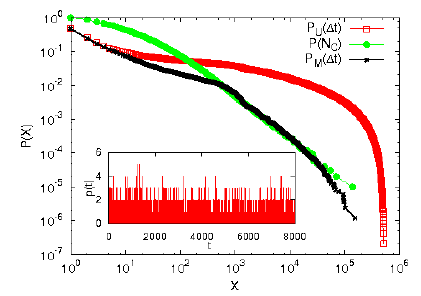}%
\hfill%
\includegraphics[width=0.495\textwidth]{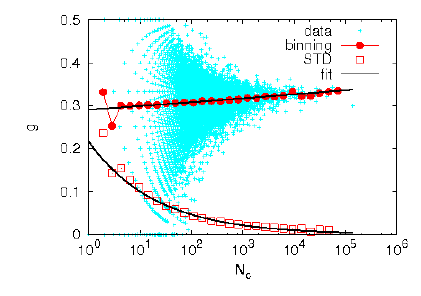}%
\caption{(Color online) Parameters of the model  $P_{\mathrm{U}}(\Delta t)$, $P_{\mathrm{M}}(\Delta t)$, $P(N_{\mathrm{c}})$ and $p(t)$ (left) and correlations between $g$ and $N_{\mathrm{c}}$ (right) inferred from the empirical data of \texttt{Ubuntu} channel. Lines show how  the mean (bullets) and the standard deviation (squares) of the data vary with $\log(N_{\mathrm{c}})$. }
\label{fig-parameters}
\end{figure}

In the spirit of individual-based modelling, the agent's $i$ participation in the chat process depends on its personal profile. Appart from the fluctuating emotional state, the agent's profile is fully defined by the following attributes \cite{BT_ABMbook}:
\begin{equation}
A\left[\text{id}, \text{type}; a_i(t),v_i(t);\mathrm{lin}_{i},\mathrm{lout}_{i}; N_{\mathrm{c}}^i,g_i; \Delta t_i; \text{status}\right] \,.
\label{eq-agent}
\end{equation}
By the agent's first appearance in the system, it receives a unique $\mathrm{id}$ and type (i.e., agent, moderator or Bot). Moreover,  to mimic heterogeneity of the real chat system, each agent gets a fixed number of messages $N_{\mathrm{c}}^i\in P(N_{\mathrm{c}})$ that it can create during the simulation time. Similarly, the agent's inclination towards direct communication $g_i$ versus posting messages on the channel, is also fixed from a distribution  $g_i\in P(g)$ by the agent's appearance. These two distributions are determined from the same empirical data to take into account  potential dependencies between these parameters, as shown in the right-hand panel, in figure~\ref{fig-parameters}. For every $N_{\mathrm{c}}$, a value of $g$ is chosen as a Gaussian random, where the mean and the standard deviation follow the dependence on $N_{\mathrm{c}}$ which is observed in the empirical data, cf. figure~\ref{fig-parameters} (right-hand). Other attributes of agents indicated in  (\ref{eq-agent}) are dynamically varying.  The agent's pattern of in-coming and out-going links, $\mathrm{lin}_{i},\mathrm{lout}_{i}$, evolves by establishing new contacts in the network. However, the agent's status (active or passive) fluctuates according to the chat process; the agent  becomes active when its delay time since it's previous performance $\Delta t_i$ expires. After each completed action, a new delay time is taken as a random number from the distribution $\Delta t_i\in P(\Delta t)$. To  statistically match the agent's profiles from a given empirical system, the distribution $P(\Delta t)$ is determined from the same empirical dataset. In figure~\ref{fig-parameters} (left-hand panel), the distribution is shown for the case of ordinary users $P_{\mathrm{U}}(\Delta t)$  and moderators $P_{\mathrm{M}}(\Delta t)$  as well as the distribution $P(N_{\mathrm{c}})$ derived from the dataset of \texttt{Ubuntu} channel. It should be stressed that, apart form statistically shorter delay times, the moderators are given an unlimited number of messages $N_{\mathrm{c}}^{\mathrm{M}}=\infty$ during the simulation time. Similarly, the Bot's capacity is unlimited. In addition, each agent whose $\mathrm{id}$ was placed on the Bot's list gets reply from the Bot without any delay. (Other types of Bots are considered in \cite{we-abm_ExpBots,we-ABMrobots}.) Note that the activity of moderators involves human characteristics, i.e., delay time, one-to-one communication as well as  fluctuating emotions under the influence of the received messages. On the other hand, the Bot's emotion is fixed during the entire simulation time; moreover, Bot simultaneously replies to all agents on its list.  Here, we consider \texttt{posBot} with a fixed positive emotion ``enthusiastic'', and \texttt{negBot} with a negative emotion ``shame'', corresponding to the coordinates ($a=0.6718,v=0.5154$) and ($a=-0.2346,v=-0.4747$), respectively, in the emotion circumplex map \cite{russell1980}. The number of messages that the Bot handles per time step vary, being regulated by the fraction $\epsilon$ of all messages of  currently active moderators. Also, for a better control of Bot's impact, in this model $g_{\mathrm{Bot}}=1$, i.e., Bot does not post messages to the channel. Note that in this model, by analogy with the empirical system, the Bot is always present; hence, it can be picked for conversation by an ordinary agent even if $\epsilon=0$.

First, the Bot is introduced and added to the empty \textit{active agents list}; the Bot's list and channel list are also started.  Subsequently, at each time step, a fluctuating number $p(t)$ of agents is added; the profile of each added agent is fixed according to equation~(\ref{eq-agent}) and its initial arousal and valence are selected as uniform random  values. The time series $p(t)$ which is used in the simulations is shown in the inset to figure~\ref{fig-parameters}; it represents the number of new arrivals in the empirical system. The time series $p(t)$ is a pink noise signal with pronounced daily fluctuations \cite{we-ABMrobots}. In this way, we incorporate the necessary daily cycles in the agents' activity. Moreover, temporal resolution of the empirical time series (in this case one minute) sets the time scale for  the simulation time step. Numerical values of the parameters  $T_0=2$min; $k=0.4$; $c_2=d_2=0.5$; $\gamma_a=\gamma_v=0.01$ are used in the simulations; for typical values of the fields, these parameters enable a larger phase space for each trajectory (see also \cite{we-ABMrobots} where different parameters are varied).

Interactions among added agents with the agents on the active list are executed according to the model rules. (The program flow is given in the Appendix.) Time step is ended by relaxing emotion variables  and decreasing delay times of all present agents and by updating the active agents list. In the simulations, we keep track of each message: its creation time, identity of the sender and the receiver (or channel), as well as its emotional content, which comprises the emotional arousal and valence of the sender at the moment of message creation.

\clearpage

\section{Occurrence of collective emotional states on chat networks\label{sec-simulations}}
\subsection{Emergence of a social network}
\begin{figure}[!h]
\hspace{10mm}
\includegraphics[width=0.4\textwidth]{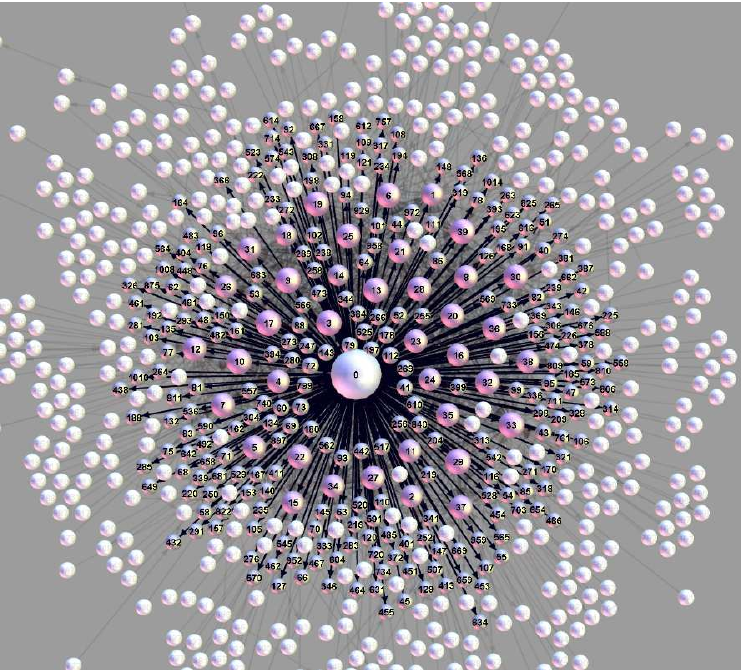}
\hfill%
\includegraphics[width=0.5\textwidth]{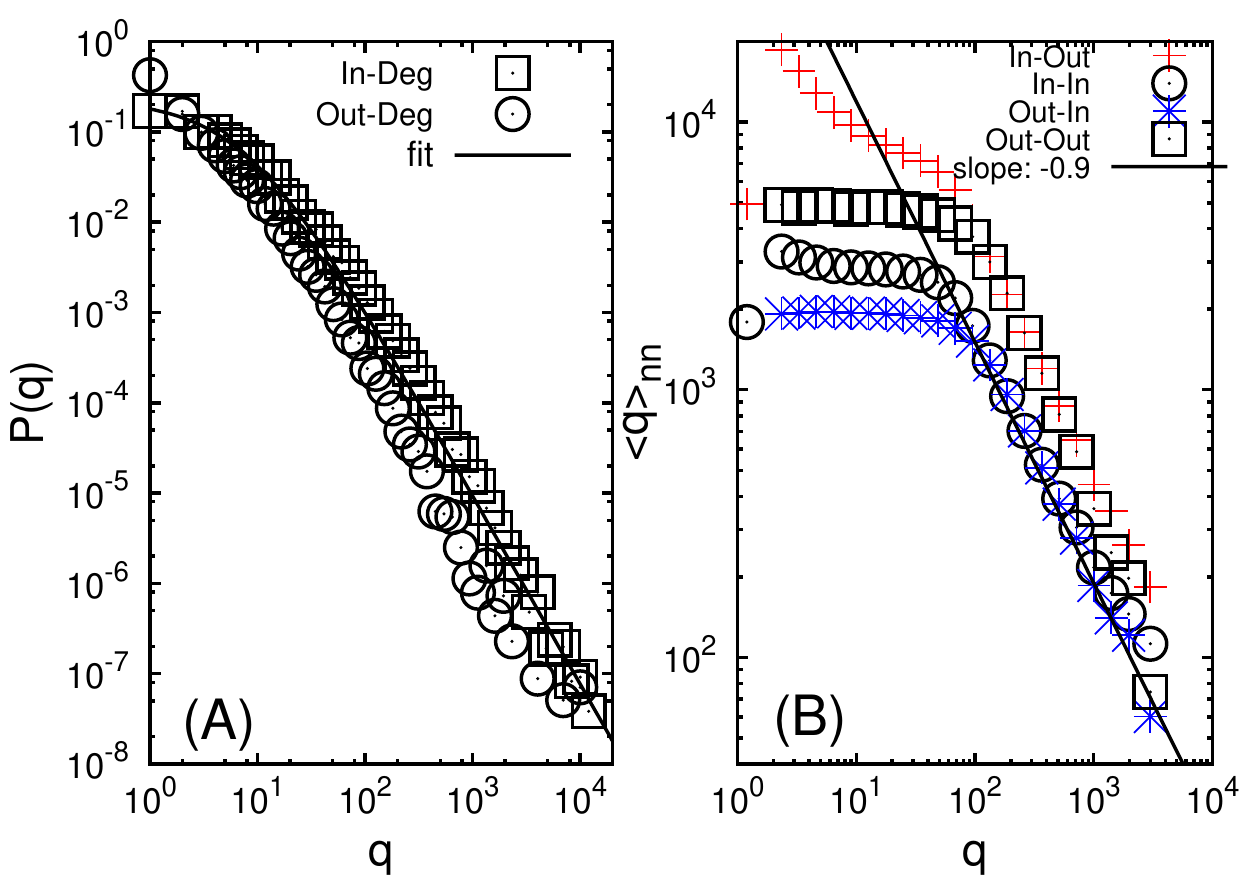}%
\hspace{10mm}
\caption{(Color online) Left: Network emerging after 2000 steps of chats with agents, moderators and \texttt{negBot} (in the center). The highlighted links indicate the subset of agent who received the Bot's messages.
 Right  (A)-(B): Topology measures of the network from large simulations. Data are from  \cite{we-Chats-chapter}.}
\label{fig-ABMchatnet}
\end{figure}

As it was mentioned in section~\ref{sec-model},  in the process of chats among agents, moderators and Bot a social network comprising all their connections evolves.
An example of such chat network appearing after 2000 simulation steps in the presence of \texttt{negBot} is displayed
in figure~\ref{fig-ABMchatnet}. As discussed in \cite{we-Chats-conference,we-Chats1s,we-entropy}, chat networks exhibit
a hierarchical structure that is strongly associated with the type
and emotional content of messages interchanged between participants. A detailed analysis of the topology of chat networks
both from empirical data and agent-based simulations within the present model is given  in \cite{we-Chats-chapter}.
It was shown that, when an emotionally neutral Bot is active and $\epsilon=0.5$ in the model, the emergent network
has statistically similar features as the network obtained from the  \texttt{Ubuntu} channel data. Specifically,
both network types are described with the power-law distributions for in-coming and out-going links with a similar
exponent $\tau_q\sim 2$ as well as disassortativity measures that are shown in figure~\ref{fig-ABMchatnet}A, B.
Moreover, the chat networks exhibit $k$-core structure with a power-law $k$-dependence \cite{we-Chats-chapter}.
The central core consists of moderators and Bot.

When the emotional Bots are active, statistically similar network topologies emerge in the case of \texttt{posBot} and \texttt{negBot}. However, their layered structure that can be distinguished according to the emotion valence of messages sent along the links is considerably different  and depends on the Bot's emotion \cite{we-Chats-chapter}. Here, we focus on quantitative analysis of underlying processes by which collective states  with positive and negative emotion emerge under the Bot's influence.

\subsection{Collective response of the network to the Bot's activity\label{sec-response}}
\begin{figure}[!h]
\centering
\begin{tabular}{cc}
\resizebox{18.8pc}{!}{\includegraphics{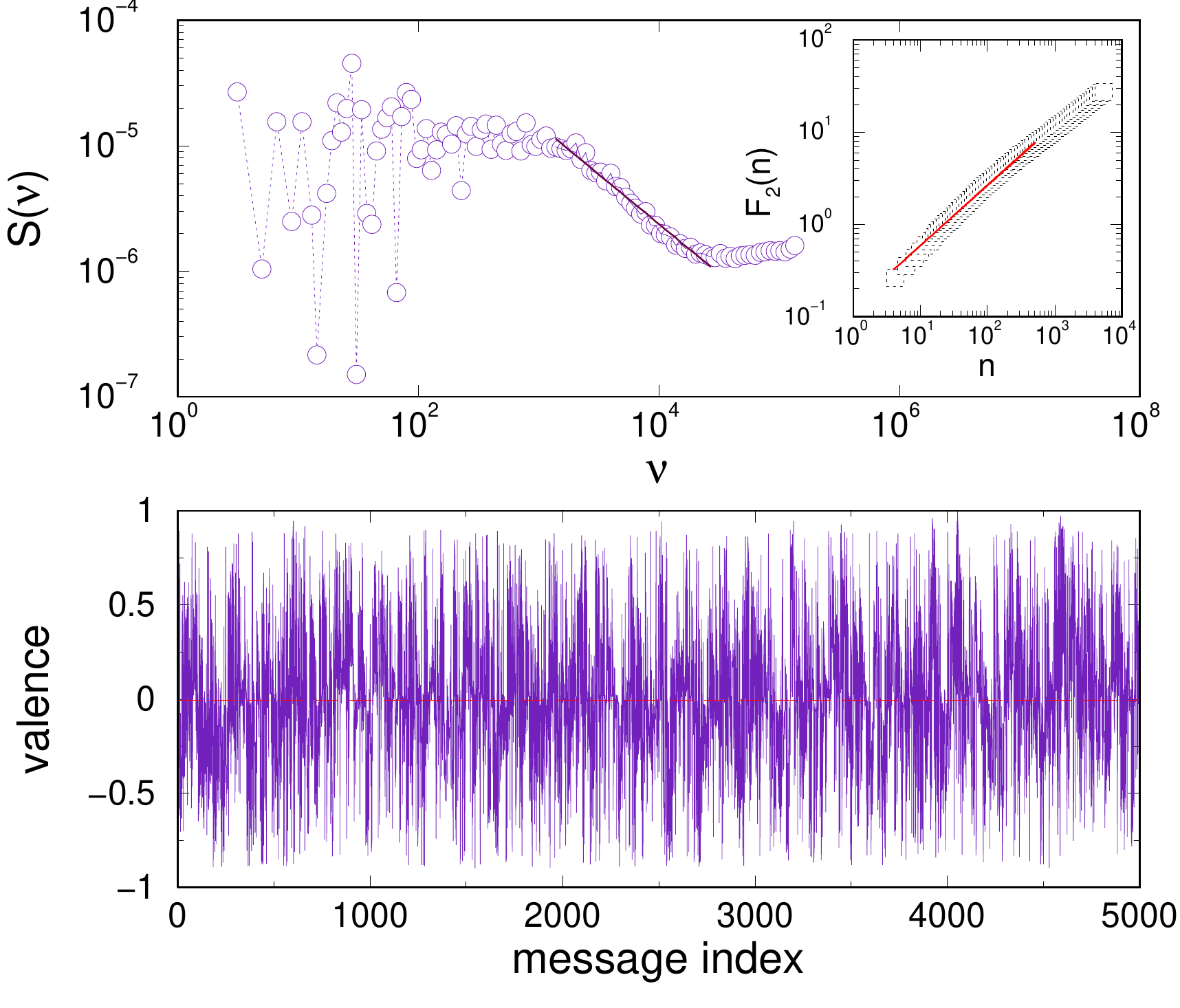}}
\\
\end{tabular}
\caption{(Color online) Bottom: Valence fluctuations in the temporal sequence of messages among agents and moderators in the  absence  of Bot. Dotted line: linear fit with $\langle v\rangle=0.00297$ and a small positive slope $8\times 10^{-8}$. For clarity, only initial 5000 messages are shown. Top: Power spectrum  and (inset) fluctuation at the time window of the length $n$ of the valence time series.
}
\label{fig-valence-noBot}
\end{figure}

In the absence of emotional Bots, the emotion polarity at the level of the entire network is well balanced. In figure~\ref{fig-valence-noBot}, we plot the actual valence in the temporal sequence of messages interchanged among the agents (including moderators) when the Bot is entirely inactive. Slightly positive emotion charge, similarly to the empirical data, can be attributed to  higher arousals for most of the positive emotions, compared with negative emotions that are involved in chats. In a limited scaling region, indicated by straight lines in figure~\ref{fig-valence-noBot}, we find the scaling exponents $H_v=0.633\pm 0.005$, $\phi _v=0.79\pm0.02$ for the fluctuations around the trend,  $F_2(n) \sim n^H$, and the power spectrum, $S(\nu)\sim \nu^{-\phi}$, respectively. However, when an emotional Bot is active, the system tends to polarize  by spreading  the Bot's emotion among  agents \cite{we-abm_ExpBots,we-entropy,we-ABMrobots}.
\begin{figure}[!h]
\centerline{\includegraphics[width=14cm]{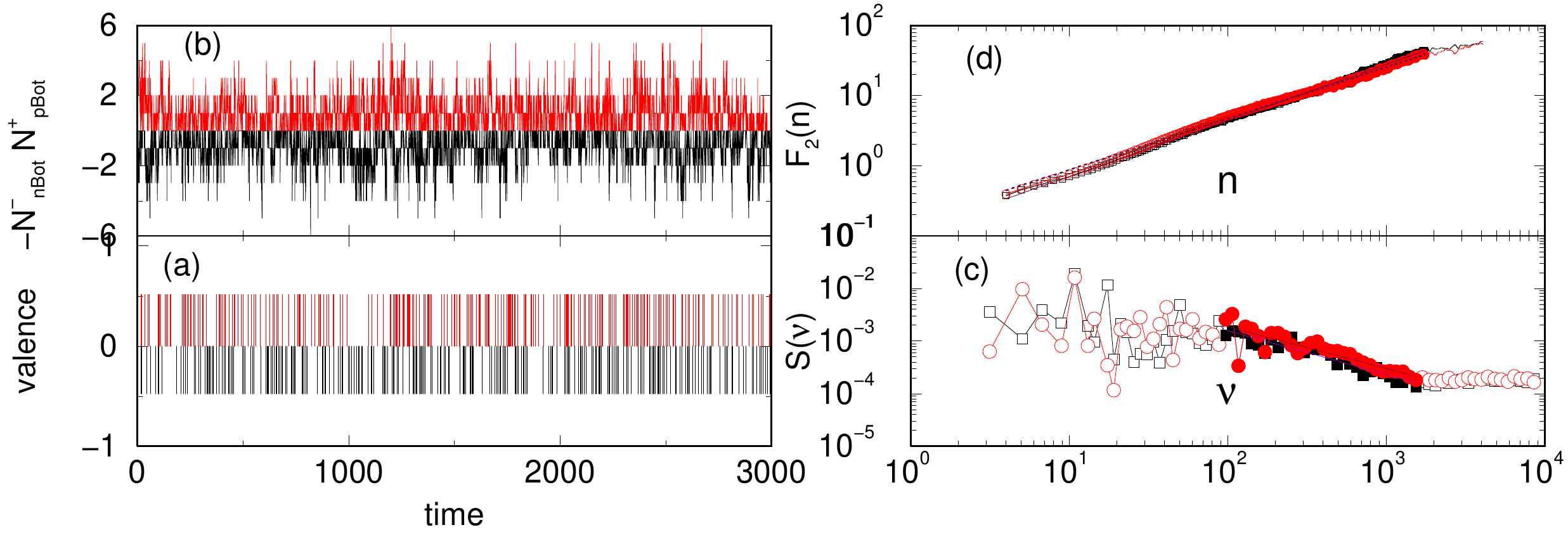}}
\caption{(Color online) (a) and (b) Sequences of the Bot's messages with fixed valence and time series of the number of matching polarity messages by agents $N^+_{\mathrm{pBot}}(t)$ and $- N^-_{\mathrm{nBot}}(t)$ for positive and negative Bot, respectively. (c) and (d) Power spectrum and fluctuations of these time series.}
\label{fig-ABMemoBots-response}
\end{figure}

On the other hand, when moderators are absent, the response of agents to the Bot's activity is depicted in figure~\ref{fig-ABMemoBots-response}. Sequences of messages by \texttt{posBot} (red) and \texttt{negBot} (black) shown in the bottom panel, induce a number of messages with the same valence polarity $N^+_{\mathrm{pBot}}(t)$ and $N^-_{\mathrm{nBot}}(t)$, respectively, by all agents in the network. In (c) and (d), the power spectrum $S(\nu)$ and the fluctuations $F_2(n)$ at a time window $n$ for these two time series are plotted. In the scaling region, indicated by straight lines,  the scaling exponents are: $H^+=0.731\pm 0.001$, $\phi ^+=0.79\pm 0.08$, for the $N^+_{\mathrm{pBot}}(t)$, and $H^-=0.799\pm 0.001$, $\phi^-=0.83\pm 0.05$, for the $N^-_{\mathrm{nBot}}(t)$ time series, respectively.

\begin{figure}[!h]
\centerline{\includegraphics[width=11cm]{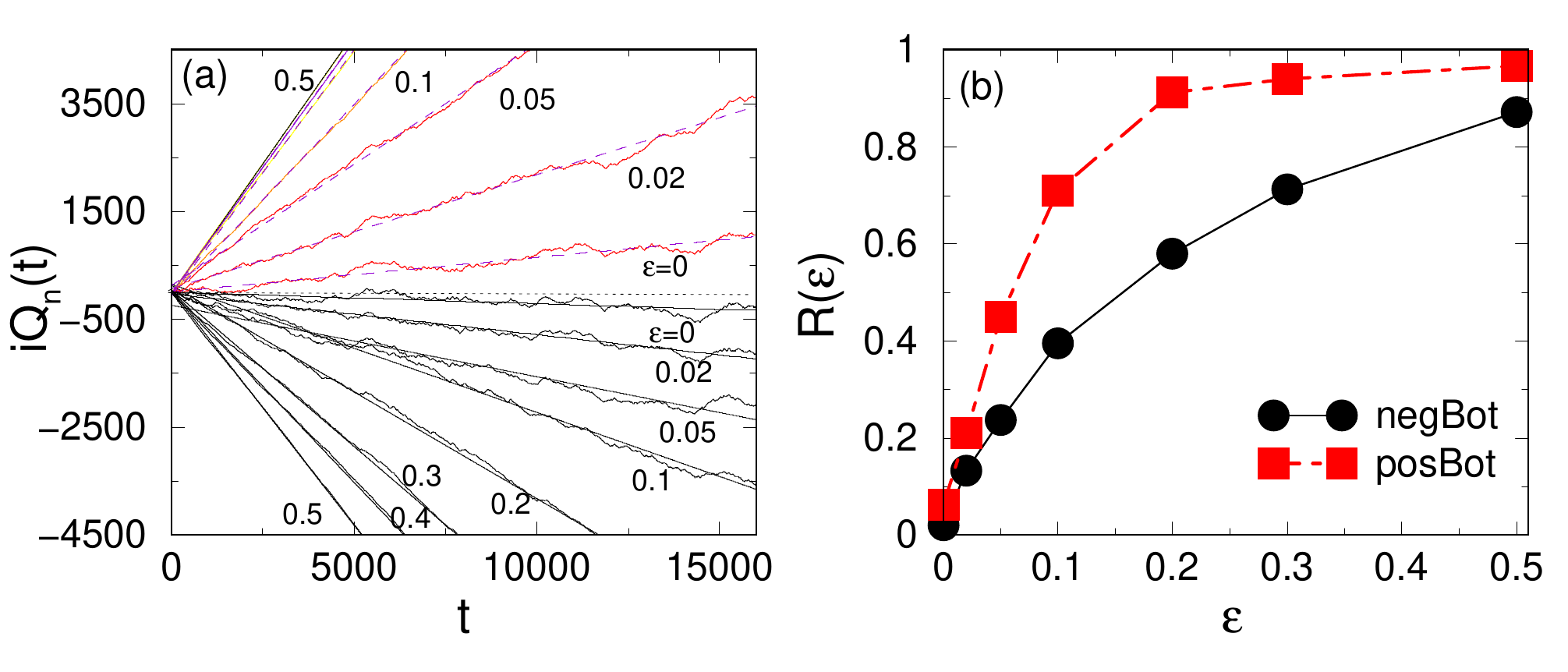}}
\caption{(Color online)  (a) Integrated normalized charge time series for positive Bot (upper half-plane) and negative Bot (lower half-plane) in the presence of moderators and varied fraction of messages  $\epsilon$. (b) Growth rate  $R$ of the normalized charge plotted against $\epsilon$. }
\label{fig-ABMemoBots-QnormEE}
\end{figure}
In the presence of Bots and moderators, the fluctuations  of the normalized charge of
emotion-con\-tained messages $Q_n(t)=\frac{N^+(t)-N^-(t)}{N^+(t)+N^-(t)}$ exhibit a trend; specifically,
$Q_n(t)$ increases towards positive values, in the presence of \texttt{posBot}, or decreases
towards negative values, when \texttt{negBot} is active. The average rate
$R=\langle\vert\frac{\rd Q_n(t)}{\rd t}\vert\rangle$ thus measures the Bot's
emotional impact on the entire system, and can serve as an \textit{order parameter}
\cite{tadic2007}. With the action of moderators, the Bot's activity level is tuned by the
fraction of messages $\epsilon$ passed to Bot.  To determine $R(\epsilon)$, the normalized
charge signal integrated until  time $t$,  $\textrm{i}Q_n(t)$,  is  plotted against  $t$ in
figure~\ref{fig-ABMemoBots-QnormEE} for different values of $\epsilon$. Upper and
lower half-planes correspond, respectively, to the situations  when \texttt{posBot}
and \texttt{negBot} were involved.  The order parameter as a function of $\epsilon$,
determined from the linear fits of these curves, is shown in figure~\ref{fig-ABMemoBots-QnormEE}~(b).
Note that the average  growth rates $R(\epsilon)$ differ  at a given  $\epsilon$ value, suggesting
potential differences in the self-organized  processes spreading impacts of the positive and negative Bot.
In the following section, we examine the fractal structure of these processes.

\section{The dynamics of chats with positive and negative emotions\label{sec-tseries}}
\begin{figure}[!h]
\centerline{\includegraphics[width=12cm]{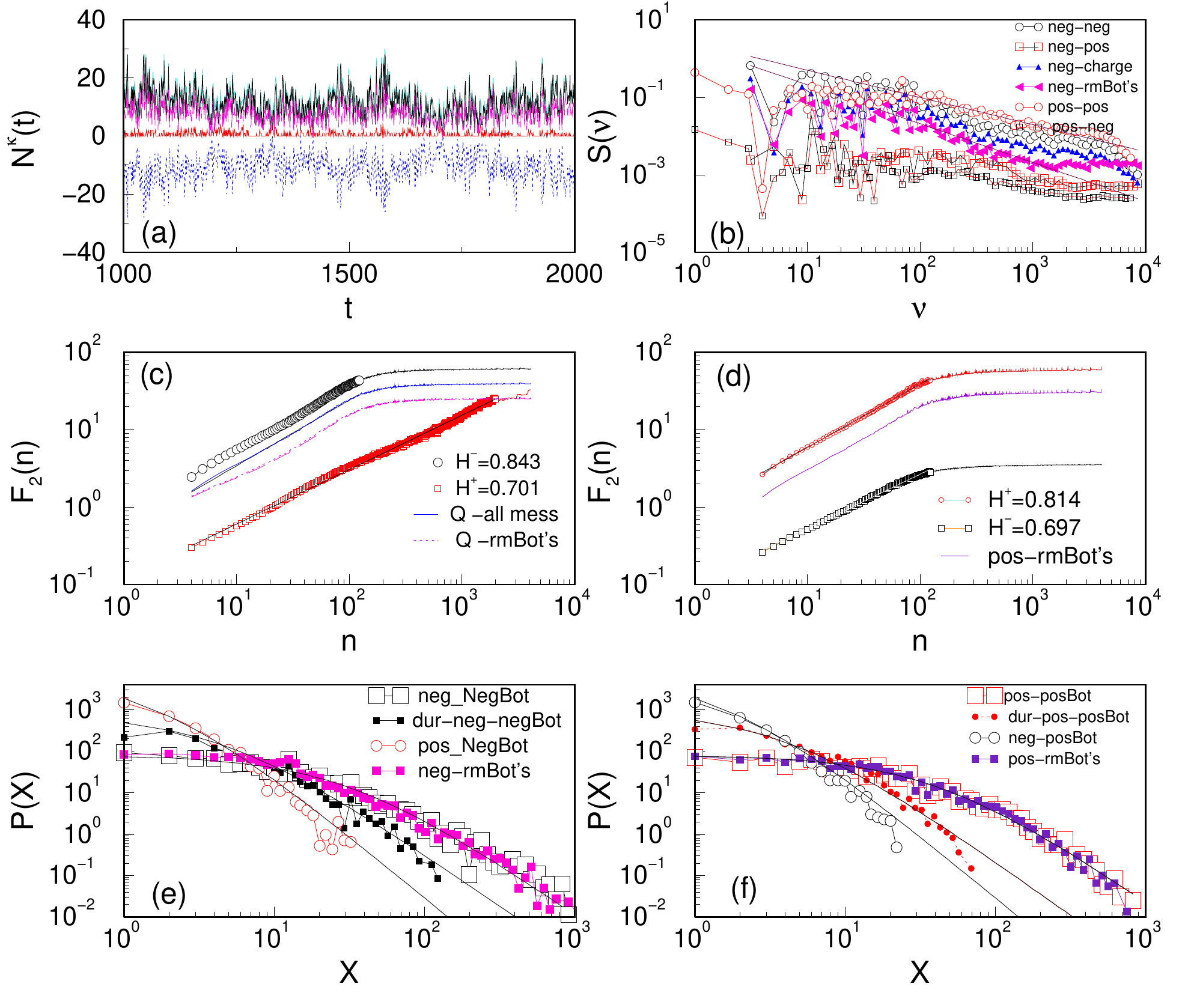}}
\caption{(Color online) Top row: (a) Time series of charge $Q(t)$ of the emotional messages (bottom, blue line), the number of messages carrying  positive (middle, red) and negative (top, black) emotion messages in the presence of \texttt{negBot}; pink line indicates the number of negative emotion messages of agents only (the Bot's messages are removed). (b) Power spectrum of these time series and time series of positive and negative messages in the case with \texttt{posBot}.
Middle: Flutuations $F_2(n)$ against time interval $n$ for the case with \texttt{negBot} (c) and \texttt{posBot} (d). Bottom: Distribution of avalanche sizes and duration of Bot-dominant emotion content and opposite to the Bot's emotion for the case of \texttt{negBot} (e) and \texttt{posBot} (f). Full lines are fits by the expression (\ref{eq-qexp}) with $\kappa =1$.
}
\label{fig-tsABMnegBot}
\end{figure}

From the simulated sequence of messages, we construct time series of interest, specifically,  time series of the number of messages carrying positive (negative) emotion valence $N^\pm(t)$, and time series of the charge of emotional messages $Q(t)=N^+(t)-N^-(t)$. Examples of such time series in the case when the \texttt{negBot} was active are shown in figure~\ref{fig-tsABMnegBot}~(a). Power spectra of these time series, as well as the corresponding time series in the presence of \texttt{posBot}, are shown in figure~\ref{fig-tsABMnegBot}~(b). Depending on the time series in question, a broad spectrum $S(\nu) \sim \nu^{-\phi}$ is found within a different range of frequencies;  the  numerical value of the exponent $\phi$ also varies, but all exponents are  in the range $\phi \lesssim 1$. Furthermore, we perform  a detrended time series analysis where, first, the dominant cycle is removed \cite{we-MySpace11}; the fluctuations around the cycle, $F_2(n)$, are plotted against the time interval $n$ in figure~\ref{fig-tsABMnegBot}~(c), (d). The indicated values of the corresponding   Hurst exponents of these time series are determined in the scaling range. These results confirm that strongly \textit{persistent} fluctuations occur in the dynamics both in the case of positive and negative collective emotion.

The physical principle behind temporal fractal structures, in various complex systems  \cite{tadic2002},
lies in a self-similarity of the underlying stochastic process, which is likely  manifested in time series
and in clustering of events, or avalanches. The occurrence of clustering in the stochastic process, on the
other hand, can be often related with nonextensive or $q$-generalized statistics \cite{tsallis2011,q-universality}.
In the present context, we consider clusters of temporally connected events that carry a specified (positive or negative)
emotion content. Using standard approaches \cite{tadic1996,mitrovic2011}, we determine the distribution of avalanche
sizes and durations; these are shown in figure~\ref{fig-tsABMnegBot}~(e), (f), corresponding to the presence of negative
and positive Bot, respectively.   The histograms are fitted by Tsallis distribution
\begin{equation}
P(X)=P_0\left[1 -(1-q)\left(\frac{X}{X_0}\right)^\kappa\right]^{1/1-q} \ ,
\label{eq-qexp}
\end{equation}
with $\kappa =1$. The non-extensivity parameter $q=1.40$ was found for the positive avalanches in the case of positive Bot and similarly, negative avalanches when negative Bot was active. A bit smaller value $q=1.33$ fits the distributions in the case of emotional valence  opposite to the Bot's emotion.

\begin{figure}[!h]
\centerline{\includegraphics[width=10cm]{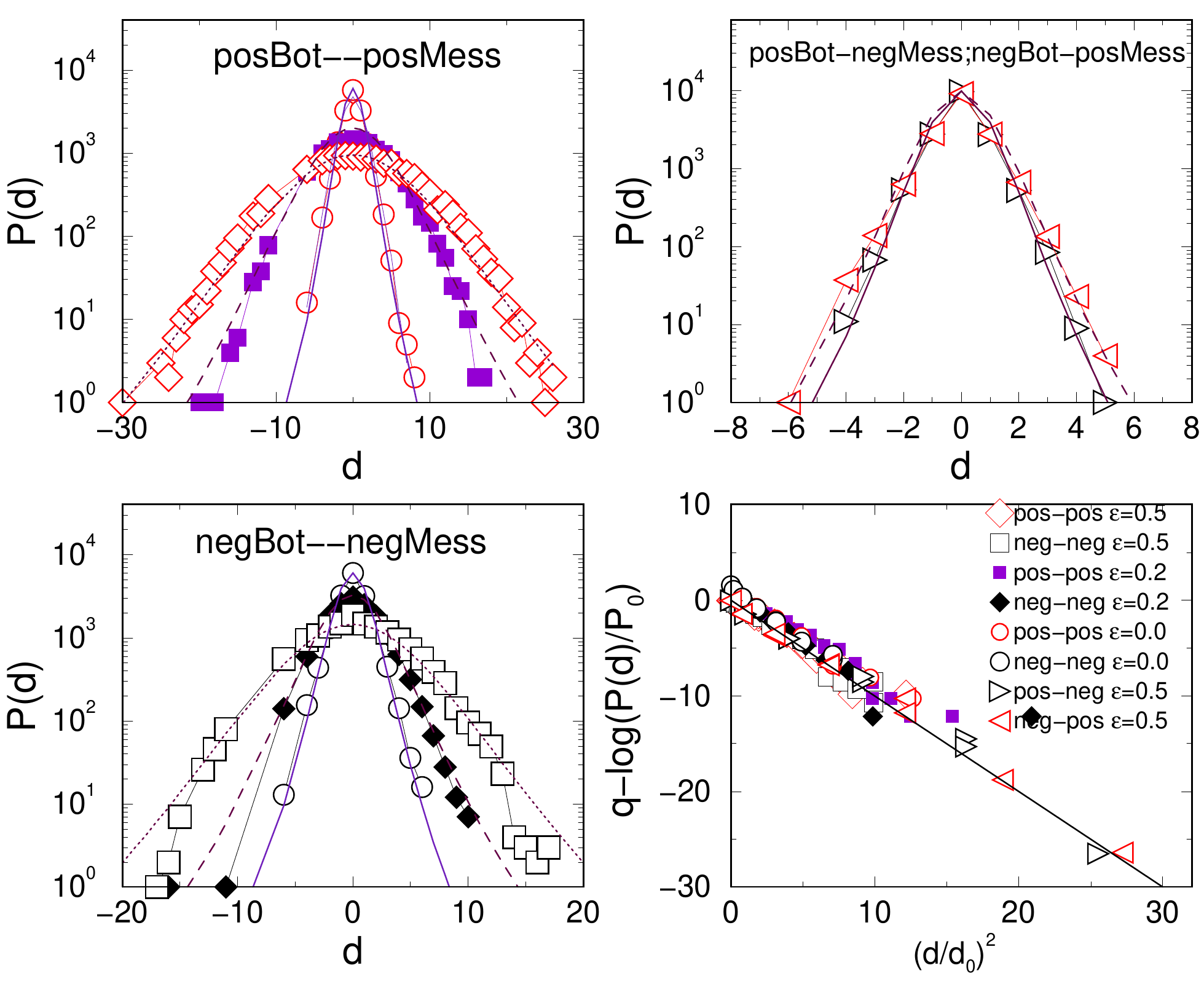}}
\caption{(Color online) Distribution of returns $d =N^\pm(t+1)-N^\pm(t)$ for the time series of positive messages $N^+(t)$ of agents in the presence of positive Bot and negative messages $N^-(t)$ in the presence of negative Bot  (left panels), and of the emotional messages with the valence opposite from the active  Bot. Lines are fits by $q$-Gaussian expression. Bottom-right-hand panel shows $q$-log plot of the distributions against the distance $d$ normalized by the respective standard deviation $d_0$. }
\label{fig-returns}
\end{figure}

The analysis of \textit{returns}, or differences between consecutive values in the considered time series,  gives further insight into the nature of relaxation processes in systems with nonextensive statistics and fractal time series. Often  $q$-Gaussian  expression (\ref{eq-qexp}) with $\kappa=2$,  is found for the distribution of returns \cite{tsallis2011,q-universality}, with $q>1$ values, strongly reflecting the strength of noise in the dynamics. For the above  time series, the distributions of returns are displayed in figure~\ref{fig-returns}. Fits  by $q$-Gaussian expression  are plotted in bottom right-hand panel  using the appropriate axes to exhibit a straight line. The related values of $q$ parameters that fit different distributions are found in a narrow range 1.20--1.25, manifesting a relative independence  of the relaxation process of the emotion contents.

\section{Conclusion\label{sec-conclusion}}
Within agent-based simulations, we have demonstrated how the presence
of emotional Bots in online chat systems can induce collective emotional
dynamics with a preset  (positive or negative) mood. In the appearance of
these collective states of agents, both the structure of the co-evolving
social network and the self-organized dynamics play their roles.
The architecture of chat networks with emotion-related layers has
been studied in \cite{we-Chats-chapter,we-entropy}. In this work,
our focus was on the collective dynamics of emotional interactions.
Considering time series of the number of messages that carry
a specified emotion content, we have analysed their fractal structure,
clustering and returns. This quantitative analysis of the corresponding
time series allowed us to uncover the role of noise (two driving modes)
in the system, on the one hand, and to relate the fractal dynamics of emotions
with the nonextensive $q$-deformed statistics, on the other hand. Our main conclusions can
be summarized as follows: (i) Owing to a higher emotional arousal that, consequently, induces
higher activity and self-organization level in the social network, positive emotions in the
presence of positive Bot propagate more efficiently than negative emotions when the negative
Bot is active. (ii)  Both positive and negative emotions, when favoured by the Bot, involve
stronger clustering and persistence in the dynamics than the  propagation of the emotion
opposite to the Bot's emotion. Hence, these characteristics of the emotion dynamics can be
attributed to the role of endogenous (emotionally) coherent noise, exerted by the Bot's activity.
(iii) External noise, represented by arrival of new agents with random emotional states, however,
almost completely controls the relaxation processes, which is manifested in the universality of
the statistics of returns.

\section*{Acknowledgements}

We are grateful for support from program P1--0044 by
the Research Agency of the Republic of Slovenia and from the
European Community's program FP7--ICT--2008--3 under grant
no 231323.     B.T. also thanks for partial support from COST
Action TD1210 KNOWeSCAPE. M.\v{S}. would like to
thank for support from projects OI171037 and III41011 of the
Republic of Serbia.

\clearpage

\appendix

\section{Model of Emotional Agents, Moderators and Bots: Program Flow}
\label{alg:ABMrobotM}

\begin{algorithmic}[1]
\STATE \textbf{INPUT:} Parameter $n_{t}$, $T_{0}$, $q$, $\epsilon$, $N_{\mathrm{M}}$, $a^{bot}$,$v^{bot}$; Distribution
$P_{M}(\Delta t)$, $P(\Delta t)$, $P(m)$, $P(g)$; Time series
$\{p(t)\}$; Start  \textit{Channel list}, \textit{Bot list} and \textit{active list} of agents;

\STATE Add Bot;
\FORALL{$1 \leqslant  t \leqslant  n_{t}$}
\FORALL{$1 \leqslant  i \leqslant  p(t)$}
\STATE Add agent; $N_{U}++$; Identify moderators; set $a_{i}\in[0,1]$ and $v_{i}\in[-1,1]$; chose
$g_i\in P(g)$ and $m_i\in P(m)$; set $\Delta t=0$; put to \textit{active list};
\ENDFOR
\STATE Calculate $h^{\mathrm{a}}_{\mathrm{mf}}(t)$ and $h^{\mathrm{v}}_{\mathrm{mf}}(t)$;
\FORALL{$i \leqslant  N_{\mathrm{U}}$ }
\STATE Calculate $h^{\mathrm{v}}_{i}(t)$ and $h^{\mathrm{a}}_{i}(t)$
\STATE Update states $v_{i}(t)$ and $a_{i}(t)$; 
\ENDFOR

\FORALL{$i \in$ \textit{active list}}
\STATE creat a message $j$; transfer $v_{i}(t)->v^{m}_{j}$ and $a_{i}(t)->a^{m}_{j}$;
\IF{$i$ is Bbot}
\STATE  send message to a selected agent from \texttt{Bot list} and the related Moderator;
\ELSE
 \IF{($\Delta t ==0\  AND  \ a_i(t) > rnd_ -next )$}
 \IF{$i$ is Moderator}
 \STATE  select an agent from Channel and with probability $\epsilon$ forward its ID to Bot; otherwise act as normal agent; 
\ELSE
    \IF{$m\leqslant  m_i$}
\STATE with probability $g_i$ select an agent from \textit{active list} and send message to it; othewise, write the message to the Channel;
\STATE $m_i++$;
    \ENDIF
  \ENDIF
 \ENDIF
\ENDIF

\ENDFOR
\STATE clear active list

\FORALL{$1 \leqslant  i \leqslant  N_{U}$}
\IF {agent $i$ was active or received a message in previos $T_{0}$ steps or $i$ is Bot}
\STATE add $i$ to \textit{active list};
\ENDIF

\IF{$(\Delta t == 0\  OR \ i \in active_-list)$}
\STATE chose new $\Delta t$ from $P(\Delta t)$ for agents and $P_{\mathrm{M}}(\Delta t)$ for moderators;
\ELSE
\STATE $\Delta t--$
\ENDIF
\STATE relax $a_i(t)$ and $v_i(t)$  with rate $\gamma$
\ENDFOR

\ENDFOR
\STATE {\bf END}
\end{algorithmic}

\newpage

\ukrainianpart

\title{Колективна динаміка емоцій у чатах з агентами, модераторами та ботами}
\author{М. Шуваков\refaddr{label1,label2}, Б. Тадічь\refaddr{label1}}

\addresses{
\addr{label1} Факультет теоретичної фізики, Інститут Йожефа Стефана, Любляна, Словенія
\addr{label2} Інститут фізики Белграду, Університет Белграду, Белград, Сербія
}

\makeukrtitle

\begin{abstract}
За допомогою агентно-орієнтованих симуляцій ми досліджуємо флуктуації колективних емоційних станів на мережі чату, в якому агенти обмінюються повідомленнями з фіксованою кількістю модераторів та емоційних ботів. Щоб спроектувати реалістичну систему чату, правила взаємодії, деякі статистичні ха\-рак\-те\-рис\-ти\-ки, а також атрибути агента базуються на емпіричному аналізі каналу чату \texttt{Ubuntu}.
У симуляціях емоції бота є фіксованими; модератори корегують рівень активності бота, направляючи до нього частину повідомлень $\epsilon$. При $\epsilon \gtrsim 0$ поступово виникає колективний емоційний стан, що співпадає з полярністю емоцій бота;  параметром порядку служить середній рівень зростання домінантного емоційного заряду.   Завдяки ефектам самоорганізації, колективна динаміка виявляється більш вираженою, коли позитивні емоції провокуються позитивним ботом, ніж при провокуванні негативних емоцій негативним ботом при однаковому $\epsilon$. Більше того, базові фрактальні процеси демонструють більшу стійкість та сильнішу кластеризацію подій у випадку, коли в системі поширена емоція із полярністю, що співпадає з емо\-цією бота, ніж коли поширюється емоція, протилежна до емоції бота. З іншого боку, динаміка загасання контролюється зовнішнім шумом; залежний неекстенсивний параметр, що визначається із статистики затримок, є віртуально незалежним від рівня активності бота та емоційного змісту.

\keywords стохастичні процеси на мережах, скейлінг у соціальній динаміці, агентно-орієнтовані симуляції, колективна емоційна поведінка
\end{abstract}

\end{document}